\documentclass[acmsmall]{acmart}

\setcopyright{rightsretained}
\acmJournal{PACMHCI}
\acmYear{2025}
\acmVolume{9} \acmNumber{1}
\acmArticle{GROUP30} \acmMonth{1}
\acmDOI{10.1145/3701209}

\makeatletter
\gdef\@copyrightpermission{
  \begin{minipage}{0.2\columnwidth}
   \href{https://creativecommons.org/licenses/by/4.0/}{\includegraphics[width=0.90\textwidth]{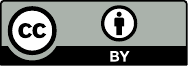}}
  \end{minipage}\hfill
  \begin{minipage}{0.8\columnwidth}
   \href{https://creativecommons.org/licenses/by/4.0/}{This work is licensed under a Creative Commons Attribution International 4.0 License.}
  \end{minipage}
  \vspace{5pt}
}
\makeatother

\begin{document}

%%
%% The "title" command has an optional parameter,
%% allowing the author to define a "short title" to be used in page headers.
\title[Online Communities for Voluntary Misinformation Response]{The Collaborative Practices and Motivations of Online Communities Dedicated to Voluntary Misinformation Response}

%%
%% The "author" command and its associated commands are used to define
%% the authors and their affiliations.
%% Of note is the shared affiliation of the first two authors, and the
%% "authornote" and "authornotemark" commands
%% used to denote shared contribution to the research.

\author{Jina Yoon}
\email{jinayoon@cs.uw.edu}
\orcid{0000-0002-2523-3922}
\affiliation{%
  \institution{University of Washington}
  % \streetaddress{}
  \city{Seattle}
  \state{WA}
  \country{USA}
  \postcode{98101}
}

\author{Shreya Sathyanarayanan}
\email{sathshr@cs.washington.edu}
\orcid{0009-0000-3251-6173}
\affiliation{%
  \institution{University of Washington}
  % \streetaddress{}
  \city{Seattle}
  \state{WA}
  \country{USA}
  \postcode{98101}
}

\author{Franziska Roesner}
\email{franzi@cs.washington.edu}
\orcid{0000-0001-8735-4810}
\affiliation{%
  \institution{University of Washington}
  % \streetaddress{}
  \city{Seattle}
  \state{WA}
  \country{USA}
  \postcode{98101}
}

\author{Amy X. Zhang}
\email{axz@cs.uw.edu}
\orcid{0000-0001-9462-9835}
\affiliation{%
  \institution{University of Washington}
  % \streetaddress{}
  \city{Seattle}
  \state{WA}
  \country{USA}
  \postcode{98101}
}

%%
%% By default, the full list of authors will be used in the page
%% headers. Often, this list is too long, and will overlap
%% other information printed in the page headers. This command allows
%% the author to define a more concise list
%% of authors' names for this purpose.
\renewcommand{\shortauthors}{Jina Yoon, Shreya Sathyanarayanan, Franziska Roesner, \& Amy X. Zhang}
%% No italics

%%
%% The abstract is a short summary of the work to be presented in the
%% article.
\begin{abstract}
Responding to misinformation online can be an exhausting and thankless task. It takes time and energy to write effective content, puts users at risk of online harassment, and strains personal relationships. Despite these challenges, there are people who voluntarily respond to misinformation online, and some have established communities on platforms such as Reddit, Discord, and X (formerly Twitter) dedicated to these efforts. In this work, we interviewed 8 people who participate in such communities to understand the type of support they receive from each other in these discussion spaces. Interviewees described that their communities helped them sustain motivation, save time, and improve their communication skills. Common practices included sharing sources and citations, providing emotional support, giving others advice, and signaling positive feedback. We present our findings as three case studies and discuss opportunities for future work to support collaborative practices in online communities dedicated to misinformation response. Our work surfaces how resource sharing, social motivation, and decentralization can make misinformation correction more sustainable, rewarding, and effective for online citizens.
\end{abstract}

%%
%% The code below is generated by the tool at http://dl.acm.org/ccs.cfm.
%% Please copy and paste the code instead of the example below.
%%


%%
%% Keywords. The author(s) should pick words that accurately describe
%% the work being presented. Separate the keywords with commas.

\received{May 2024}
\received[revised]{August 2024}
\received[accepted]{October 2024}

%%
%% This command processes the author and affiliation and title
%% information and builds the first part of the formatted document.
\maketitle
\section{Introduction}
Responding to misinformation is not an easy task. It requires significant time, energy, and trust to develop high-quality and source-backed content~\cite{wilner2023time, malhotra2023user}. Being actively involved in such efforts also puts people in positions of high visibility and, consequently, high risk for harassment ~\cite{schafer2023towards}. Confronting someone about a misleading post can also strain or ruin relationships, especially with friends or family~\cite{scott2023figured, feng2022investigating}. Past work has also shown that corrections may not actually be very effective~\cite{ecker2022psychological} and can even potentially backfire~\cite{mosleh2021downstream}. Results of social corrections are heavily dependent on the receiver of the correction~\cite{martel202maybe}, and this lack of control over positive outcomes can lead to feelings of burnout and pointlessness for people who confront misinformation regularly, such as health professionals ~\cite{bautista2021physicians}.

Despite these challenges, there are still many people who \textit{voluntarily} respond to misinformation online. Some have established communities on discussion-based platforms such as Reddit, Discord, and X to connect with others who engage in similar efforts including social correction, crowdsourced fact-checking, or peer victim support. In this work, we interviewed eight online citizens from such communities about how they interact with each other and present our findings as three case studies grouped by each online community's primary discussion topic and purpose. Case 1 focuses on a Discord server where regular contributors of Community Notes on X (formerly known as Birdwatch on Twitter) collaborated to produce politically balanced fact-checking notes. Case 2 consists of interviews with from members of r/QAnonCasualties and relevant communities where users discuss how conspiracy theories have affected themselves or their loved ones. Case 3 focuses on r/vaxxhappened, a subreddit where users repost screenshots of misleading information from external sources to debunk or vent about. Our research questions were the following:

\begin{itemize}
    \item RQ1: What \textbf{motivates} people to voluntarily respond to misinformation?
    \item RQ2: What is the \textbf{purpose of community} for responders?
    \item RQ3: What are the \textbf{factors that shape the collaborative practices} within these communities?
    \item RQ4: What can we learn from existing communities and apply to \textbf{future community-oriented misinformation interventions}?
    \end{itemize}

Our interviews revealed the many collaborative practices that occur in these online communities and the ways peer interactions influenced individual motivations. In our discussion, we highlight three main themes participants relayed in their experiences. First, the resources and information shared by other members in these online communities saved users significant time and energy. Additionally, the social support found in these online communities helped to sustain motivation and provide valuable feedback. Finally, we observed that the decentralized nature of these online communities could produce more localized and engaging communications. We offer these considerations and opportunities for future work, and we emphasize the need for further research given our limited sample of participants from Western cultures.

In summary, this work explores the collaborative practices of online communities dedicated to voluntary misinformation by interviewing eight participants about their interactions with peers in their Reddit and Discord communities. We present our findings as three case studies with a focus on users' motivations. Our discussion explores opportunities to support and leverage their collaborative practices and underscores the need for research on global contexts. This empirical study adds to existing work on community-based approaches to combating misinformation by surfacing the collaborative practices of online communities dedicated to misinformation response and the potential ways they might increase or sustain motivation for more online citizens.

\section{Related Work}
\subsection{Social Corrections of Misinformation and the Effects of Perceived Norms}
A large proportion of misinformation intervention research tends to focus on individual and platform-level interventions. On the individual level, prior misinformation intervention research has been conducted on topics such as developing a standardized set of credibility indicators for identifying credible content in articles~\cite{zhang2018structured}. On the platform level, misinformation intervention work has focused on the design of ``lightweight interventions'' that ``nudge'' the user to evaluate the accuracy and credibility of content before sharing on social media~\cite{jahanbakhsh2021exploring}. Additional research has been conducted about the ways in which individuals interact with and investigate misinformation on their social media feeds~\cite{geeng2020fake}. There has also been related work about source-related misinformation interventions such as how media provenance can impact a user’s perception about the trust and accuracy of social media content. Provenance is the information about the origins of a piece of digital media and any modifications made~\cite{feng2023examining} Other source-related mechanisms include developing checklists, toolkits, and the use of expert sources for individuals to evaluate online misinformation~\cite{heurur2022comparative}. However, as Aghajari et al. emphasize, these types of individual and platform-based interventions place responsibility on the user and make the assumption that users are rational actors~\cite{aghajari2023reviewing}. These approaches also fail to account for the social contexts in which individuals interact with misinformation and do not necessarily consider systemic impacts. Thus, more recent research in misinformation intervention has advocated for more socially embedded strategies such as social corrections and perceived norms~\cite{aghajari2023reviewing}.

Social corrections are when people directly reply to or confront the individual who posted the misleading information. However, these are very difficult: they often lead to highly emotionally charged conversations, which can put pressure and tension on relationships, especially when the interaction occurs between people who are close at a personal level such as family and friends~\cite{scott2023figured}. It can also be very tiring because it takes time to write responses with research-backed sources~\cite{bautista2021healthcare, bautista2021physicians, wilner2023time}. Engaging in misinformation response in a social setting can also expose people to harms like harassment and bullying ~\cite{schafer2023towards}. In addition to these risks, it also often does not feel rewarding since people are unlikely to change their mind immediately after being socially corrected~\cite{ecker2022psychological}. Thus, if we want to encourage social corrections, we should simultaneously look at ways to alleviate these burdens and make the experience less harmful for people, which is a core motivation for this work.

Another important social factor that has been studied is the effect of perceived norms on behaviors toward misinformation. Perceived norms have been shown to strongly influence people’s behavior and attitudes toward believing in and sharing information. These can be leveraged to encourage positive actions and combat misinformation~\cite{gimpel2021effectiveness}. However, these same mechanisms can reinforce and perpetuate false beliefs or rumors, especially in closed networks~\cite{difonzo2013rumor}. Other examples of community characteristics that can influence a community member’s response to misinformation include the user’s interaction patterns within a community, perceived norms around content produced and shared, network structures, and overall perspectives on different issues~\cite{aghajari2023reviewing}.

\subsection{Participatory Collaboration in Crowd-Based Knowledge Production}
The motivations of voluntary participation in online communities of crowd-base knowledge production has been well-documented in CSCW literature about Wikipedia contributors~\cite{kittur2008harnessing, kittur2010beyond, zhang2017crowd, zhu2013effects}, Reddit moderators\cite{seering2023moderates, squirrell2019platform, seering2019moderator}, and open-source software developers~\cite{vasilescu2013stackoverflow, vasilescu2014social}. Some have utilized social capital theory to explain the motivations of sustained voluntary participation in such communities~\cite{qiu2019going, nemoto2011social}. In the present study, we aim to similarly understand the motivations of those who participate in another form of crowd-based knowledge production: crowdsourced fact-checking. Crowdsourced fact checking has a long history rooted in rumors and crisis response in which online citizens working together to dispel rumors in order to assist during emergent situations, such as the aftermath of the 2010 Haiti Earthquake ~\cite{starbird2011voluntweeters}. Though these volunteers usually have positive intentions, online citizens unfortunately can make mistakes when attempting to verify unsubstantiated information as was surfaced after the 2013 Boston Marathon bombing~\cite{starbird2014rumors}.

More recently, new social computing systems have been developed to formalize and support crowds in fact-checking endeavors. In 2020*, X launched Community Notes~\footnote{https://help.twitter.com/en/using-x/community-notes} (formerly known as Twitter and Birdwatch) and was one of the first prominent platforms that supported crowdsourced fact-checking. Community Notes allowed users to create community-driven ``notes'' using a collaborative approach to provide more informative context to Tweets and address misinformation. The notes that contributors create are displayed on Tweets based on whether enough community members rate the note as helpful to, which helps ensure that diverse perspectives are considered~\cite{lorenz2022birdwatch, x2024algorithm}. Existing research on Community Notes has focused mostly on the labor, value, reliability, and effectiveness of the system and its data~\cite{jones2022misleading, drolsbach2023diffusion}. Our work differs in that we are more interested in the motivations and collaborative practices of Community Notes contributors. We hypothesize that crowd fact-checkers may be more motivated by notions of civic duty than contributors in more subject-oriented crowd knowledge work like those on Wikipedia or Reddit.

Lastly, another area of research we draw upon is the collaborative practices of journalists and professional fact-checkers to verify information on the job~\cite{mcclure2020misinformation, micallef2022true}. Journalists and communicators rely on their professional communities and networks to validate information on breaking news together. Similarly, professional fact checkers develop pipelines of practices to overcome challenges they face when it comes to improving effectiveness, efficiency, scale, and reach~\cite{sehat2024misinformation}. We examine whether these individual-level and community-level practices also arise among voluntary fact-checkers and similarly explore potential sociotechnical solutions that could assist those who engage in this work.

\section{Methods}
\subsection{Recruitment}
We broadly defined our criteria for communities of interest as ``online discussion forums where people discuss how to correct, respond to, or heal from the harms of misinformation''. Examples of communities we reached out to included subreddits related to medical information, public health, climate change and action, vaccination information, the COVID-19 pandemic, and QAnon or other conspiracy theories. For Reddit communities, our procedure was to first message the moderators for permission to post the study recruitment information in the subreddit and/or directly message active posters and commenters in the community within the last month. Response rates were very low with this method, but we ultimately recruited three Reddit users this way. We also sought to interview people who contributed to Community Notes on X. We first recruited participants by searching for \#CommunityNotes on the platform and manually DMing users who openly revealed being an active contributor of the feature. This method had very low response rates and we only recruited one user this way. To find more Community Notes contributors, we leveraged network and snowball sampling. 

\subsection{Participants}
We interviewed a total of three users from Reddit and five from X for a total of eight participants who engage in online communities dedicated to voluntary misinformation response. All participants were 18 years or older and based in the USA except for P3 and P6 who were located in other Western countries. To protect participant privacy, we have chosen to intentionally omit or obfuscate further details about individuals and the communities they participate in.

\subsection{Limitations}
A significant limitation of our work is that all of our participants were from North American and European regions. Given that social practices and perceived norms vary greatly between cultures, especially when it comes to correcting misinformation ~\cite{badrinathan2024corrections}, we underscore this limitation and emphasize the need for further work to explore our research questions in global contexts. Additionally, given our small sample size, this work is not to be considered as a representative study. We present our findings as initial investigations in online community-based approaches for supporting voluntary misinformation response.

\subsection{Interviews}
We held semi-structured, 60 minute interviews over a platform of the participant’s choice including Zoom video calls, phone calls, Reddit live chat, or Twitter DMs. As a thank you for their time and participation, participants were offered a digital gift card of \$15 USD within 4-6 weeks of concluding the interview. Two participants refused the gift card. The interview protocol was designed to answer our research questions about collaborative practices and motivations. We asked about the interviewee’s participation in their relevant communities, the platforms they contributed on, their motivations for contributing, experiences with internal and external harassment about the topic, and how they gauge the success of their contributions. Some questions were added, revised, or removed between interviews as we reached saturation. As a result of slow recruitment and low response rates, interviews were held sporadically over the course of 6 months. 

\subsection{Analysis}
To analyze our interview data, the first author reviewed their notes and annotated transcripts between each interview. Upon completing eight interviews, the first author conducted a first round of open coding~\cite{saldana2021coding} on Taguette, an open-source qualitative interview analysis tool. They then developed an initial codebook to discuss with the other authors. After merging several codes, the first and second authors used ATLAS.ti to perform Braun \& Clarke’s reflexive thematic analysis~\cite{braun2023doing} on the same randomly selected transcript. They then compared their coding results to iterate on the codebook to develop a second version. Finally, the first author qualitatively coded four transcripts, and the second author qualitatively coded the other four. They discussed interesting themes on a weekly basis over six weeks, and the result of this analysis is the foundation for our Findings. Following this process, we noticed patterns in the desired platform features that interviewees described. To examine these more systematically, we performed structural coding to segment and categorize quotes about specific features that were mentioned by participants (e.g., Discord text channels, notifications, upvotes). We then affinity diagrammed these on a Miro board to observe clusters of features that served similar purposes across different platforms. This analysis served as the basis for our discussion.

\section{Results}
\subsection{Case 1: Collaborative fact-checking practices among early adopters of Community Notes}
Community Notes is a feature on X that allows users to add additional context (called a \textit{note}) to any post. If a note receives enough ratings from other X users to be determined as \textit{helpful}, the note becomes publicly shown on the post for all other users. Helpfulness ratings are based on a complex bridge-based ranking algorithm that is designed to find agreement between users with a wide variety of perspectives~\cite{x2024algorithm}. Despite its crowd-based inner workings, Community Notes does not natively offer a place for users to communicate with each other outside of the content of the notes themselves and the helpfulness rating system. Thus, in this case study, we studied the interactions between contributors who were invited to participate in the official Discord server for early adopters of Community Notes. According to P2, a former X employee, the development team launched this small Discord for users to provide feedback and feature requests for Community Notes. Over time, the Discord unintentionally evolved into an organic community where contributors would collaborate synchronously to produce high-quality notes and other fact-checking content.

\subsubsection{Political diversity by design}
Much like the ethos of Community Notes itself, the Community Notes Discord centered the values of political balance and diverse perspectives. According to P2, members were intentionally curated through an invite-only process where the team worked with machine learning engineers to identify users from diverse political backgrounds based on their Community Notes history. At the time of our interviews, the feature and Discord server were limited to users from the US only.

The Birdwatch Discord server therefore consisted of a broad range of users from across the US political spectrum. However, according to participants, the Discord's population seemed to skew left, likely as a result of the platform's general demographic makeup at the time. One member, P5, explained that this skew was their primary motivation for participating on Community Notes and the Discord. They identified themselves as conservative and had a sizable following on X, where they often posted political news and opinions:
\begin{quote}
``My main motivation for using [Community Notes] was I didn’t want [it] to be dominated by one political ideology … I'm right-leaning and online tends to be fairly left-leaning, so [I just wanted to be] there to bring that balance to a new program.'' —P5
\end{quote}

\subsubsection{Moderating internal conflicts}
As a result of the political diversity of the group, intra-community conflict was not uncommon. Debates about information credibility were common and sometimes got heated. However, these disagreements were not viewed as inherently problematic. As P2 described, it was expected and even encouraged given that the mission of Community Notes was to mediate differing views through discourse and crowdsourcing.
\begin{quote}
``This is a space of deliberation … [In this community], you have the opportunity to engage in conversation with someone in a moderated environment.  The whole [Community Notes] team was in it … so if things start to get heated, someone [would] moderate it. So of course there were times where [things] did get heated and then we gave like a cooldown for some people.'' —P2
\end{quote}
Moderation efforts in this community were thus largely focused on resolving internal conflicts. Additionally, since the community was private and invitation-only, participants did not relay experiences about harassment from external sources.

\subsubsection{Community Notes, community values}
In the early days of the Community Notes Discord, members frequently debated what constituted a ``helpful'' or high quality note. According to P2, the Community Notes team intentionally did not define the term so as to encourage users to establish these values as a community. Eventually, early adopters in the Discord devised a set of guidelines that were integrated into the platform:
\begin{quote}
``Early on, we had a lot of discussions on how to write notes … because most people don't know how to write quality material. So we'd [go] back and forth, `Hey, what does a good note look like? … Does a typo or poor grammar make a bad note? … Write intelligent notes that's neutral in a neutral voice.' So now, [Community Notes] gives a notification when you sign up about how to write a good note.” —P5
\end{quote}

Relatedly, members later requested that moderators create another text channel to discuss and monitor the opposite: low quality notes. In this space, members raised awareness about Community Notes users who consistently misused the feature. For example, notes that consisted of opinions or editorial statements rather than objective and neutral content were disfavored:  
\begin{quote}
``\#bad-faith-alerts [was a channel] where some people were keeping track of who the really egregious [Community Notes] users were. [For example,] any Elon Musk tweet you can see 10 different notes from people just saying mean things [that were] not really productive for the conversation.'' —P5
\end{quote}

\subsubsection{Resource-sharing and sensemaking}
Over time, many other collaborative practices and systems were established in the Community Notes Discord, and most of them were initiated by members organically. For example, \#sources-and-citations was a text channel where users shared credible sources or contested the reliability of others such as Wikipedia. Other channels that contributors requested included \#rapid-response and \#breaking-news which were spaces where users collaborated to quickly draft notes in response to what they called ``high velocity'' tweets.
\begin{quote}
``We requested a [channel] called \#rapid-response [where we put things that] needed to get through the system really fast because fake news can spread pretty quickly on Twitter … We [also] had an area called \#breaking-news [to discuss] current events … that's where a lot of notes came from.'' —P5
\end{quote}
P4 recounted the \#templates channel where they would drop simple and reusable messages for others to craft quick and effective notes at scale:
\begin{quote}
``I did share some [templates] to the Discord, but I'm not sure if anyone used mine. I made them because there were a lot of repeated talking points or false claims that required retyping of the same information, and it made it faster to write notes at a large scale … I was kind of competitive about [my score] though, so I'm not sure how much I shared.'' ---P4
\end{quote}
Both P4 and P5 also created something they called ``watchlists'', a tool for keeping an eye on updates from high-profile users. P1 expressed wishing there was a native feature on X to curate a feed of highly visible users to write notes on. Additionally, over time, specific community members became known for certain subject matter expertise, such as one member who had a research background in natural disasters. P8 recalled that this user was frequently tagged in conversations when others needed help verifying sources for issues like earthquakes and hurricanes.

\subsubsection{Impact through scale}
Another core motivation that many members of the Community Notes Discord server shared was a desire to have impact through scale. For some users, this manifested in being strongly motivated by metrics, scoreboards, and competitive systems, such as for P4:
\begin{quote}
``I had one of the highest helpful rating counts according to my own parsing of their public data downloads. Unfortunately they've since switched to different metrics and all of my numbers were reset, and I've been inactive lately, so I no longer have the impressive stats ... But yeah, if there’s a score on anything im probably going to grind it.'' —P4 
\end{quote}

P5 was not as interested in the scores displayed within the community, but they did care about the number of views that notes garnered from general X users:
\begin{quote}
``We wouldn't care if some anonymous person with no followers is tweeting crazy stuff, because nobody's gonna see that. We care about the amount of eyes that are on the tweet itself ... I couldn't care [less about] who wrote [a note]. I just care that it's a good quality note.'' ---P5
\end{quote}

\subsection{Case 2: Peer support and advice for QAnonCasualties}
r/QAnonCasualties is a Reddit community dedicated to being a safe space for people who have been adversely impacted by QAnon, a large conspiracy theory group in the US~\cite{wendling2020qanon}. Since its inception, the movement has grown and now has many relevant sibling communities across platforms including Discord and TikTok. These communities offer peer support for people who suffer heavy emotional, social, and political tolls as a result of being close to someone who believes in conspiracy theories~\cite{moskalenko2023secondhand}. %P3 was a general member of one such community, and P6 was a moderator from a European community similar to r/QAC.

\subsubsection{A safe space for healing and recovery}
% \paragraph{Creating an environment of nonjudgmental compassion and support}
r/QAnonCasualties and its related communities are spaces designed to offer a place for people with deep (and often painful) histories with the topic to heal and recover. P6, for example, recounted their own difficult past:
\begin{quote}
``For three years, [I had] been an informal conspiracy theorist. And just about two years ago, I just promptly started this subreddit. The intention was just to give people a place to communicate and to connect … the most [common topic in the community] is family. [People talk about their] aunt, mother, kid, husband; I would say 80\% of the content is like that.'' —P6 
~\end{quote}
Because the community involves such sensitive topics, moderators of r/QAnonCasualties and many of its related communities work hard to maintain the environment by enforcing strict policies against outsiders and harassers. 

\subsubsection{Venting about similar experiences}
The most common types of posts in these communities were ``venting'' posts. Members would express their frustrations and hardships and peers often responded with relatable stories and emotional validation. 
\begin{quote}
``You can really see how some people learn from the interactions and then stick around and try to lift up others who come in new and the sense of connection saying, `I'm in the same shit basically.' It's uplifting for everybody. If you have a community where [they] know the same shit you're experiencing … [people will respond]. `I know it's hard, but I have the same at home times 10.''' — P6 
\end{quote}
Notably, these communities were usually small, niche, and not intended for a wide audience. Traffic in these spaces was not very high, but the few success stories that were posted were very motivating and inspiring. P6 recalled, 
\begin{quote}
''One of the very first prominent posts was about a young man whose mother hid his vaccination papers from him. And he was asking [for advice, like] ‘I want vaccination, but my mother hides the documents from me.’ … a few months later, he posted that he got his vaccinations, moved out, and he's feeling much better now. [These success posts do not happen] very often, but I really love that the number of posts like that is not zero.'' —P6
\end{quote}

\subsubsection{Providing feedback and encouragement}
Another way members of this community supported each other was by providing encouragement and validation. P6 explained that they had recently started creating YouTube content for this topic and that the support of others in the community was key to sustaining their motivation:
\begin{quote}
``Sometimes the work is not a joy, but the results are really a blessing. You sit hours of hours and make a YouTube video and then you sit there and think that is the lowest quality thing everybody has ever produced, a nobody will watch that. And then you'll upload it and the response is overwhelmingly positive ... I often had the feedback that now they are able to laugh about this serious issue because they were really depressed.'' —P6
\end{quote}
They further elaborated that this type of feedback was meaningful because they lacked the feedback loop to know whether if their work had a tangible impact:
\begin{quote}
``If someone sees your content and they take the info from that video and then go to their QAnon aunt, there's no way to know that this [had any impact] ... I think [I have had an impact], but I cannot be 100\% sure ... I'm sure I have invoked a lot of emotions in a lot of different people. But I have not the experience of someone who says, `Yeah, you helped me to discuss better with my Nazi aunt.' So I don't know yet.'' ---P6
\end{quote}

\subsubsection{Creating local resources and live events}
% \subsubsection{Live group voice and video calls to practice communication skills}
P6 explained that their European version of r/QAnonCasualties community was very helpful for discussing conspiracy theory harms specific to their region since medical, social, and psychological resources often varied between countries and languages:
\begin{quote}
``Mostly I try to point the people that we have a wiki with service centers and psychology resources like ... How would you describe that in English? A public service [where you can call and explain something like], `I have this trouble in my family and my husband is a conspiracy theorist, what should I do, etcetera.''' ---P6
\end{quote}

P6 also participated in a relevant sibling community on Discord where users practiced conversation techniques commonly used to help individuals out of conspiracy theories. Examples include encouraging reflection, questioning information-seeking methods, and creating open dialogues. People in this community actively engaged in peer learning and mentorship through events like weekly practice sessions for ``logic interviews''. In these live Discord calls, users took turns roleplaying as an interviewer or interviewee. The interviewee typically acted as someone with strongly held beliefs about a controversial topic like religion, health, or politics, and the interviewer would ask questions to understand the interviewee.
\begin{quote}
``The word `classes' is a bit misleading; it's just a loose training group of people who meet on Sundays. We have some silly [small talk] for 10 minutes and then, hey, if somebody wants to train, we ask, `Do you have a claim to discuss?' And then everyone shuts off their camera except for the two participating [in the logic interview]. After the interview, other people give feedback like what questions could have been asked better.'' ---P6
\end{quote}

Another common activity included moderated debates about contentious topics such as philosophy or religion. Lastly, another unique practice in this group was a yearly meetup where local members of the community would gather in person to film and edit a communication technique video based on a topic of interest specific to their region.
\begin{quote}
``We have a yearly community meeting where we try to make YouTube videos by sitting in a park and inviting people to engage in such conversations.'' ---P6
\end{quote}

\subsection{Case 3: Snark, sarcasm, and science on r/vaxxhappened}
r/vaxxhappened is a subreddit where members repost media found elsewhere, usually created by vaccine opposers on external social media platforms, and collaboratively debunk or criticize it. Examples include screenshots from Facebook Groups, Instagram, or X, as well as articles from smaller news websites. Communities like these are often referred to as ``snark'' communities on Reddit, or spaces where members use irony or sarcasm to ridicule content from elsewhere. Although snark is often associated with derogatory and destructive intentions, some have argued that it serves productive functions similar to the role of gossip in other social contexts~\cite{tsiveriotis2017everything}. 

\subsubsection{Actively seeking misinformation}
P3 was a member of r/vaxxhappened and shared that one of the ways they discovered content to repost was by joining and silently observing communities of vaccine opposers from other platforms. They explained that they participated in these practices out of personal interest:
\begin{quote}
``I do a lot of personal research (i.e., not for any institution, school, or employer) on the QAnon conspiracy theory [by joining] a ton of pro-QAanon Telegram groups, gab communities ... I'm not pro-QAnon though, I just find their conspiracy really interesting to follow and I've been following it from the beginning. It's wild.'' —P3
\end{quote}

When asked about whether they engage with QAnon users in these spaces, they replied that they usually did not unless they felt someone was in danger:
\begin{quote}
``I won't reach out to them on their own turf [since] I don't think I'll get through to anyone if I go to their community and try to discuss my views, even if it's rationally. If I see someone posting something dangerous though I'll speak up, it just depends where.'' ---P3
~\end{quote}

\subsubsection{Curating quality content}
In r/vaxxhappened, community members are typically aligned in their views about scientific and medical research. To maintain this culture, the moderation policies are strict about outsiders. ``Antivaxxers'' are banned without warning if they appear to be spreading harmful views or theories. Another policy enforced is that ``low effort'' posts are removed without warning in order to ensure high quality and humorous content for the community:
\begin{quote}
    ``Low effort submissions are discouraged: Participate at your own risk. Crappy memes and image macros, reposts and shitposts may be removed.'' ---Community Rule \#5
\end{quote}

\subsubsection{Memes with a purpose}
Despite the ``snarky'' culture of the community, P3 said that they found r/vaxxhappened to be a valuable resource. Members frequently engaged in collaborative information sharing in the process of debunking reposts:
\begin{quote}
   ''I do believe r/vaxxhappened is a useful community. It allows people to share information that is oftentimes harmful to let everyone know what some people are currently believing about a certain topic (like COVID vaccines, flu shots, newborn vaccines etc). In the comments of some posts you'll often see the accurate information on that given topic with links to research papers and information.'' ---P3
\end{quote}

Similar to instances in Case 2, people would sometimes reply to posts with relevant personal stories to either vent and validate, or to provide advice and share successful strategies. P3 recalled reading about others' past experiences and that these helped improve their own communication skills:
\begin{quote}
``Oftentimes, posters who have deradicalized people they know describe what worked for them and what didn't. How they approached them, what words triggered them, etc. Learning that helps me tailor how I address people online when I'm trying to get through to someone who believes in something similar.'' ---P3
\end{quote}   

\section{Discussion}
\subsection{Online communities save volunteer responders time, effort, and energy}
In a 2023 study, Wilner et al. surfaced that information professionals who promote digital literacy face significant challenges due to resource limitations, especially in urgent and time-sensitive situations~\cite{wilner2023time}. One of the ways that their participants coped with this pressure was to build upon existing resources from other information professionals in online forums such as lesson plans and instructional materials. Participants in our study engaged in similar practices like content co-creation, resource sharing, and peer learning saved members time, energy, and effort. For example, in Case 1, members of the Community Notes Discord innovated various optimizations such as the \#sources-and-citations, \#templates, and \#rapid-response text channels. By combining their information and resources, they reduced the amount of time it took to detect, research, and respond to ``high velocity'' tweets. The group's collective knowledge also benefited from contributions of users with niche subject matter expertise.

Although other communities in our study were not as explicitly dedicated to fact-checking, they still developed similar practices that improved misinformation response efficiency. In Case 2, users in r/QAnonCasualties and related communities shared success stories and communication tips. P6's weekly Discord calls were another example in which members shared their time and resources to improve each other's skills. Additionally, when r/vaxxhappened users debunked misinformation through Reddit post replies, their comments became reusable archives of sources and citations. Future work could explore developing sociotechnical systems specifically designed for these communities of practice to save voluntary misinformation responders significant time, energy, and effort. For example, commonly upvoted sources, tips, and citations could be organized and pinned for future access and reuse.

\subsection{Online communities may reduce burnout and increase motivation for responders}
Correcting misinformation can feel pointless or futile~\cite{gurgun2023challenging}, and it can even harm personal relationships~\cite{feng2022investigating}. Past work has shown that social norms can counter these potential negative consequences and increase likelihood of engaging in corrections~\cite{aghajari2023reviewing}. Our findings point to similar opportunities to encourage voluntary misinformation response through social motivation in online communities. For example, P6 described that the nature of their work debunking conspiracy theories was ``not a joy'', but the encouragement and humor they found through their community kept them going. Relatedly, the comedic relief that users produced in r/vaxxhappened was central to the community's growth and success. Simply having a place to meet others with similar values, goals, and motivations may help sustain otherwise thankless efforts.

In addition to providing emotional support, peer interactions can also help responders gauge the success of their communications. Bautista et al. found that many health professionals lose motivation to engage in social media and online communications due to the lack of measurable positive outcomes~\cite{bautista2021physicians}. P6 similarly described that they have no way of knowing how many people have productive conversations with loved ones as a result of their YouTube content. P4 from the Community Notes Discord stopped sharing their templates with others because they did not know if anyone was using them. Future work could investigate how to capture and transform peer support in these communities into visible metrics. Implementing such systems could make responders feel more appreciated for their voluntary labor, similar to how some Reddit moderators favored Moderator Appreciation Days or profile badges~\cite{dosono2019moderation, cai2021moderation}. Additionally, our participants all had intrinsic motivations that compelled them to join such communities in the first place. Studying ways to sustain their motivations could unlock opportunities to promote voluntary misinformation correction as a social norm for general populations beyond already-interested users. 

\subsection{Online communities offer localized \& decentralized approaches to misinformation response}
Chen et al.’s recent work on community-based approches to combating misinformation surfaced the importance of working with local community leaders such as faith leaders, youth group leaders, and bilingual leaders of immigrant communities~\cite{chen2024circle}. Collaborating with these trusted messengers is especially effective for reaching marginalized groups since misinformation can be very sensitive, racialized, and politicized. In our study, we observed how P6's European ``spinoff'' of r/QAnonCasualties facilitated resources and events local to their region. The community was created because the original r/QAnonCasualties community was for English speakers in the US, which was not always accessible or applicable for international users. For example, the conspiracy theory groups active in their country went by a different name, and they frequently engaged in code switching throughout their interview to describe community phenomena. Future work should examine the potential of online communities for localized voluntary misinformation response, especially since our study is limited to users from Western cultures.

Another potential benefit of online communities for misinformation response is that their decentralized nature might reach larger and more peripheral audiences. Starbird et al. found that in \textit{dis}information campaigns, online communities drew informal participation from a wide range of users from forum moderators to activist grandmothers~\cite{starbird2019disinformation}. These  ``citizen marketers'' engaged in a ``self-sustaining web of interdependent collaborative practices'' similar to what is found in online fan communities. Our findings in Case 3, r/vaxxhappened, surfaced a similar ecosystem of media production and dissemination. Users created memes and entertaining content, and these artifacts would carry forward messages in unpredictable ways. P3, our main participant of Case 3, was not from the US but still participated in the US-centric community because they found the collaborative practices interesting and fun. This could also explain why the subreddit has over 350k members despite its niche topic and strict moderation policies. Future approaches to community-based misinformation response should explore how to blend top-down, institutional support with grassroots, voluntary community-based action. This is especially important since members like P3 today frequently expose themselves to misinformation and potential harassment when seeking content to repost. More organized efforts could protect users from these harms and scale these communities' impact.

\section{Conclusion}
In this work, we presented findings from eight interviews as three case studies of online communities dedicated to voluntary misinformation response: the Community Notes Discord, r/QAnonCasualties and spinoff communities, and r/vaxxhappened. We identified various collaborative practices and influences on participants' motivation. In our discussion, we explored how these communities can save responders' resources, sustain motivation, and produce more effective content. We offer potential future directions of research and emphasize the need to study these phenomena in global contexts. In summary, this work explores opportunities for online community-based approaches for combating misinformation and seeks to support and expand upon the collaborative practices of volunteer misinformation responders.
%%
%% The acknowledgments section is defined using the "acks" environment
%% (and NOT an unnumbered section). This ensures the proper
%% identification of the section in the article metadata, and the
%% consistent spelling of the heading.
\begin{acks}
The authors thank Kevin Feng, Connie Moon Sehat, and the Hacks/Hackers team for their extensive feedback and editing support. They would also like to thank all interviewees for taking the time to share their experiences with them and the volunteer moderators of their respective online communities for making this work possible. This material is based upon work supported by the NSF CSGrad4US Fellowship under Grant No. G-1A-016. Any opinions, findings, and conclusions or recommendations expressed in this material are those of the authors and do not necessarily reflect the views of the National Science Foundation.
\end{acks}

%%
%% The next two lines define the bibliography style to be used, and
%% the bibliography file.
\bibliographystyle{ACM-Reference-Format}
\bibliography{refs}

\end{document}